\documentclass[12pt]{article}
\usepackage{amssymb,amsmath,epsfig,graphicx}
\begin{document}
\title{\bf Dissipative Cylindrical Collapse of a Charged Anisotropic Fluid in $f(R)$ Gravity}

\author{M. Farasat Shamir\thanks{farasat.shamir@nu.edu.pk} and
M. Atif Fayyaz \thanks{atif{\_}fayyaz@yahoo.com}\\\\
Department of Sciences and Humanities, \\National University of
Computer and Emerging Sciences,\\ Lahore Campus, Pakistan.}

\date{}

\maketitle
\begin{abstract}
This paper is devoted to investigate the cylindrical collapse of an anisotropic fluid in $f(R)$ gravity. For this purpose, the viscous charged anisotropic fluid dissipating energy with heat flow and shear is assumed. We use the perturbation scheme to develop the dynamical equations for the variables that ultimately lead to the disturbance of the physical variables and the Starobinksy like $f(R)$ model chosen. The evolution of the matter variables is discussed with the help of these equations. It can be concluded that the range of dynamic instabilities depends on the field strength, density distribution, pressure and the curvature term of the $f(R)$ model. We find that our results of Newtonian and post-Newtonian regimes reduce asymptotically to general relativity solutions in the limiting case.

\end{abstract}


\section{Introduction}

It has been depicted that anisotropy can play a vital role in changing evolution and various physical properties of stellar objects. It is shown that maximal surface redshift of isotropic and anisotropic stars may differ from each other \cite{nai1}. Astrophysicists are giving their utmost attention towards the origin of anisotropy specially within the stellar course. Effects and origin of local anisotropy has been reviewed \cite{san1, chan6}. Herrera et al. \cite{her4} gave a detailed study about spherically symmetric dissipative anisotropic fluids. All the physical procedures and process that may cause deviation from isotropy can be inquired both in high and low density regimes. At higher densities, effects of short range repulsions have dominance and lessen the pionic effects giving the multiple values for pressure \cite{har1}. Matrinez \cite{mar1} pointed out that nuclear densities are effected by viscosity of neutrino which ultimately can change the gravitational collapse of viscous stars. Spacetime is divided by radiating star into two different regions, i.e the interior region and the exterior region. Junctions conditions for matching interior region and exterior region of spacetime must be satisfied. These conditions were first formed by Santos \cite{san2}. Later on various models of shear free radiative collapse were studied in great details\cite{kol1}-\cite{gov2}. Moreover pressure, anisotropy \cite{san4} and electromagnetic field \cite{mah2} were also included to generalize junction condition.

The gravitational collapse of a star is one of interesting topic of discussions. Astrophysicists aim to construct realistic models for collapsing star with radiations. Complicated non linear equations are needed to be solved to get realistic models. Most of the previous works only consider shear free motion of the fluid \cite{de1}-\cite{wag1}. It would be interesting to include shear in the analysis as it provides interesting results in the gravitational collapse of a star \cite{nog1}-\cite{josh1}. Collapse and expansion of relativistic anisotropic self gravitating source has been discussed by Abbas \cite{abb2}. Pressure of the star at the start of the collapse is believed to be isotropic but presence of shear made the pressure anisotropic. It has been shown that the presence of shear is the reason for anisotropy in self gravitating bodies \cite{chan9}. Dynamical instability of spherically symmetric, adiabatic, non-adiabatic and shearing viscous fluid were examined in \cite{her1}-\cite{her3}.

In recent past, prospects of collapse with reference to dark sources are worked out. In particular, the topic has gained much popularity in the context of modified theories of gravity \cite{sha3}-\cite{sha6}. Sharif and Rizwana \cite{sha1} discussed the dynamics of spherically symmetric gravitational collapse in $f(R)$ gravity. The same authors \cite{sha2} studied the effects of dark energy on the dynamics of the collapsing fluid in metric $f(R)$ gravity and it was concluded that dark energy affects the mass of the collapsing matter. The possibility of forming anisotropic compact stars with cylindrical symmetry has been discussed by Abbas et al. \cite{abb}. Abbas and Sarwar \cite{abb1} studied the stability of the expansion free self gravitating source in the frame work of Einstein Gauss-Bonnet gravity.

The most widely studied modified theory in the last decade is $f(R)$ gravity \cite{tho41}-\cite{kau41}. Some interesting works have also been published in the context of $f(R,T)$ gravity \cite{ifra40}-\cite{zub41}. Thus it seems interesting to investigate the issue in
$f(R)$ theory of gravity. In this paper, we are interested to discuss cylindrical collapse of a charged anisotropic fluid in $f(R)$ theory. The paper can be divided mainly in four sections. Section $2$ provides a brief introduction of modified field equations in the presence of charge and cylindrical symmetry. The perturbation scheme and the discussions of instability range for Newtonian and post Newtonian regimes are given in section $3$. Final section gives a brief summary and conclusion of the work.

\section{Modified Field Equations}

Here we consider timelike three dimensional collapsing cylindrical boundary surface $\sum$ filled with an anisotropic charged fluid, that divides four dimensional line element into two different manifolds $V^{+}$ and $V^{-}$.
For the interior $V^{-}$ space-time we take the general non-static cylindrically symmetric metric in the co-moving coordinates given by \cite{Guha}
\begin{equation}\label{1}
d{s}^{2}_{-}=-A^{2}(t,r)dt^{2}+B^{2}(t,r)d{r}^{2}+C^{2}(t,r)(d\theta^{2}+dz^{2}).
\end{equation}
The line element for exterior region $V^{+}$ is considered as
\begin{equation}
d{s}^{2}_{+}=-\left(1-\frac{2M}{r}+\frac{Q^{2}}{r^{2}}\right)d{v}^{2}-2drd{v} +r^{2}(d\theta^{2}+dz^{2}),
\end{equation}
where $M$ and $Q$ represent the mass and charge of the fluid respectively.
The action for $f(R)$ gravity is
\begin{equation}\label{2}
S=\frac{1}{2}\int d^{4}x \sqrt{-g}\left(\frac{f(R)}{\kappa}-\frac{F}{2\pi}\right),
\end{equation}
where $\kappa$ is a coupling constant, $F=\frac{1}{4}F^{{u} {v}}F_{{u} {v}}$ is Maxwell invariant with $F_{{u} {v}}=\phi _{{v} ,{u}}-\phi _{{u} ,{v}}$ representing electromagnetic field tensor and $\phi _{{u}}=\phi(t,r)\delta ^{0} _{{u}}$ denotes the four potential.
The field equations for $f(R)$ gravity are
\begin{equation}\label{2}
f_{R} R_{{u}{v}}-\frac{1}{2}f(R)g_{{u}{v}}-\bigtriangledown_{{u}}\bigtriangledown_{{v}} f_{R}+g_{{u}{v}} \Box f_{R} = \kappa (T_{{u}{v}}+E_{{u}{v}}).
\end{equation}
Here $f_{R}=\frac{df(R)}{dR}$, $\bigtriangledown_{{u}}$ denotes covariant derivative, $\Box = \bigtriangledown^{{u}}\bigtriangledown_{{u}}$, $T_{{u} {v}}$ is minimally coupled stress-energy tensor and $E_{{u} {v}}$ is electromagnetic tensor defined as
\begin{equation}\label{7}
E_{{u}{v}}= \frac{1}{4\pi}(-F^{\omega}_{{u}}F_{{v}\omega}+ \frac{1}{4}F^{\omega x}F_{\omega x}g_{{u}{v}}).
\end{equation}
Field Eqs (\ref{2}) can be expressed in a more familiar form as
\begin{equation}\label{3}
G_{{u}{v}}= \frac{\kappa}{f_{R}}[L_{{u}{v}}],
\end{equation}
where
\begin{equation}
L_{{u}{v}}=T^{(D)}_{{u}{v}}+T_{{u}{v}}+E_{{u}{v}},
\end{equation}
and effective stress-energy tensor $T^{(D)}_{{u}{v}}$ is defined as
\begin{equation}\label{4}
T^{(D)}_{{u}{v}}= \frac{1}{\kappa}\Bigl[\frac{f(R)-Rf_{R}}{2} g_{{u}{v}}+\bigtriangledown_{{u}}\bigtriangledown_{{v}}f_{R}-g_{{u}{v}}\Box f_{R}\Bigr].
\end{equation}
Here we assume that the matter is adiabatic and anisotropic which represents  dissipative collapse with heat flux and is given as \cite{Guha}
\begin{equation}\label{5}
T_{{u}{v}}=(\rho + p_{\perp})V_{{u}}V_{{v}}+p_{\perp}g_{{u}{v}}+(p_{r}-p_{\perp})\chi _{{u}} \chi _{{v}}+q_{{u}}V_{{v}}+q_{{v}}V_{{u}}+l_{{u}}l_{{v}}-2\eta \sigma _{{u} {v}},
\end{equation}
where $\rho$, $p_{r}$, $p_{\perp}$, $V_{{u}}$, $q_{{u}}$, $l_{{u}}$, $\chi _{{u}}$ and $\eta$ represent the energy density, the radial pressure, the tangential pressure, velocity four vector, heat flux, null vector, radial four vector and the coefficient of shearing viscosity respectively.
The shear tensor $\sigma _{{u} {v}}$ is defined as
\begin{equation}\label{6}
\sigma _{{u} {v}}= \frac{1}{2}[(V_{{u} ;{v}} +V_{{v} ;{u}})+(a_{{u}}V_{{v}}+a_{{v}}V_{{u}})]-\frac{1}{3}\Theta(g_{{u} {v}}+V_{{u}}V_{{v}}),
\end{equation}
where the $4$-acceleration $a_{{u}}$ and the $\Theta$ are given by
\begin{equation}
a_{{u}}=V_{{u}; {v}}V^{{v}},~~~~~~~~~~~\Theta = V^{{u}}_{;{u}}.
\end{equation}
Using the Bianchi identities, it follows that
\begin{eqnarray}\label{BianchiId}
{L^{{u}{v}}}_{;{v}} V_{{u}}=0 ,\qquad  {L^{{u}{v}}}_{;{v}} X_{{u}}=0.
\end{eqnarray}
Using co-moving coordinates, we consider the following constraints:
\begin{equation}
V^{{u}}=A^{-1}\delta ^{{u}} _{0},~~~ q^{{u}}=qB^{-1}\delta ^{{u}} _{1},~~~ \chi ^{{u}}=B^{-1}\delta ^{{u}} _{1},~~~ l^{{u}}=A^{-1}\delta ^{{u}} _{0}+B^{-1}\delta ^{{u}} _{1},
\end{equation}
such that
\begin{equation}
V^{{u}}V_{{u}}=-1,~~~ \chi ^{{u}} \chi _{{u}}=1,~~~ \chi ^{{u}}V_{{u}}=0,~~~ q^{{u}}V_{{u}}=0,~~~ l^{{u}}V_{{u}}=-1,~~~ l^{{u}}l_{{u}}=0.
\end{equation}
The Maxwell's equations are given by
\begin{equation}\label{maxwell}
{F^{{u}{v}}}_{;{v}}=4\pi j^{{u}},\qquad F_{{u}{v};\omega}=0,
\end{equation}
%with $$F_{01}=-F_{10}=-\frac{\partial \varphi}{\partial r}.$$
where $j^{{u}}\equiv{\mu} (t,r)V^{{u}}$ is four current, and ${\mu}$ represents charge density.
Using Eqs. (\ref{1}) and (\ref{maxwell}), we obtain
\begin{equation}\label{8}
\frac{\partial^{2}\varphi}{\partial r^{2}}-\left(\frac{A'}{A} + \frac{B'}{B} - \frac{2C'}{C}\right)\frac{\partial\varphi}{\partial r}= 4\pi{\mu} AB^{2},
\end{equation}
and
\begin{equation}
\frac{\partial^{2}\varphi}{\partial t \partial r}-\left(\frac{\dot{A}}{A} + \frac{\dot{B}}{B} - \frac{2 \dot{C}}{C}\right)\frac{\partial\varphi}{\partial r}= 0,
\end{equation}
where dot and prime represents derivative with respect to $t$ and $r$ respectively.
Integrating Eq.(\ref{8}), we get
\begin{equation}
\frac{\partial\varphi}{\partial r}=\frac{qBA}{C^{2}},
\end{equation}
such that electric field intensity $E$ within a region of radius $r$ and total charge $q$ are related by the following equations
\begin{equation}
q= \int {\mu} BC^{2} dr ,\qquad   E= \frac{q}{4\pi C^{2}}.
\end{equation}
For the interior spacetime (\ref{1}), field equations (\ref{3}) takes the form
\begin{eqnarray}\label{9}\nonumber
&&\frac{\dot{C}}{C}\left(\frac{2\dot{B}}{B}+\frac{\dot{C}}{C}\right)+ \left(\frac{A}{B}\right)^{2}\left(-\frac{2C''}{C}+\frac{C'}{C}\left(\frac{2B'}{B}-\frac{C'}{C}\right)\right)=
\frac{A^{2}}{f_{R}}\bigg[\kappa\left(\rho +1+2\pi E^{2}\right)\\&&+\frac{f-Rf_{R}}{2}+\frac{f_{R}''}{B^{2}} -\frac{\dot{f_{R}}}{A^{2}}\left(\frac{\dot{B}}{B}+\frac{2\dot{C}}{C}\right)-\frac{f_{R}'}{B^{2}}\left(\frac{B'}{B}-\frac{2C'}{C}\right)\bigg],
\end{eqnarray}
\begin{eqnarray}\label{10}
2\left(\frac{\dot{C}'}{C}-\frac{\dot{B}C'}{BC}-\frac{\dot{C}A'}{CA}\right)=\frac{1}{f_{R}}\bigg[\kappa\left(q+1\right)AB-\left(\dot{f_{R}'}-\frac{A'}{A}\dot{f_{R}}-\frac{\dot{B}}{B}f_{R}'\right)\bigg],
\end{eqnarray}
%\begin{eqnarray}\label{11}\nonumber
%&&-\left(\frac{B}{A}\right)^{2}\left(\frac{2\ddot{C}}{C}+\left(\frac{\dot{C}}{C}\right)^{2}-
%\frac{2\dot{A}\dot{C}}{AC}\right)+\left(\frac{C'}{C}\right)^{2}+\frac{2A'C'}{AC}=\frac{B^{2}}{f_{R}}\bigg[\kappa (p_{r}+1-\frac{4}{\sqrt{3}}\eta \sigma\\&&
%\nonumber-2\pi E^{2})
%-\frac{f-Rf_{R}}{2}+\frac{\ddot{f_{R}}}{A^{2}}-\frac{\dot{f_{R}}}{A^{2}}\left(\frac{\dot{A}}{A}-\frac{2\dot{C}}{C}\right)
%-\frac{f_{R}'}{B^{2}}\left(\frac{B'}{B}+\frac{2C'}{C}\right)\bigg],
%\end{eqnarray}
%Or
\begin{eqnarray}\label{12}\nonumber
-\left(\frac{B}{A}\right)^{2}\left(\frac{2\ddot{C}}{C}+\left(\frac{\dot{C}}{C}\right)^{2}-\frac{2\dot{A}\dot{C}}{AC}\right)+\left(\frac{C'}{C}\right)^{2}+\frac{2A'C'}{AC}=\frac{B^{2}}{f_{R}}\bigg[\kappa (p_{reff}+1 \\
-2\pi E^{2})-\frac{f-Rf_{R}}{2}+\frac{\ddot{f_{R}}}{A^{2}}-\frac{\dot{f_{R}}}{A^{2}}\left(\frac{\dot{A}}{A}-\frac{2\dot{C}}{C}\right)
-\frac{f_{R}'}{B^{2}}\left(\frac{B'}{B}+\frac{2C'}{C}\right)\bigg],
\end{eqnarray}
%\begin{eqnarray}\label{13}\nonumber
%G_{22}=-\left(\frac{C}{A}\right)^{2}\Bigg[\frac{\ddot{B}}{B}+\frac{\ddot{C}}{C}-\frac{\dot{A}}{A}\left(\frac{\dot{B}}{B}+\frac{\dot{C}}{C}\right)+\frac{\dot{B}\dot{C}}{BC}\Bigg]+\left(\frac{C}{B}\right)^{2}\Bigg[\frac{A''}{A}+\frac{C''}{C}
%-\frac{A'}{A}\left(\frac{B'}{B}-\frac{C'}{C}\right)\\ \nonumber
%-\frac{B'C'}{BC}\Bigg]=\frac{C^{2}}{f_{R}}\bigg[\kappa\left(p_{\perp}+\frac{2}{\sqrt{3}}\eta \sigma+2\pi E^{2}\right)-\frac{f-Rf_{R}}{2}+\frac{\ddot{f_{R}}}{A^{2}} -\frac{f_{R}''}{B^{2}}-\frac{\dot{f_{R}}}{A^{2}}\left(\frac{\dot{A}}{A}-\frac{\dot{B}}{B}-\frac{\dot{C}}{C}\right)\\
%\nonumber
%-\frac{f_{R}'}{B^{2}}\left(\frac{A'}{A}-\frac{B'}{B}+\frac{C'}{C}\right)\bigg],
%\end{eqnarray}
%Or
\begin{eqnarray}\label{14}\nonumber
&&-\left(\frac{C}{A}\right)^{2}\Bigg[\frac{\ddot{B}}{B}+\frac{\ddot{C}}{C}-\frac{\dot{A}}{A}\left(\frac{\dot{B}}{B}+\frac{\dot{C}}{C}\right)+\frac{\dot{B}\dot{C}}{BC}\Bigg]+\left(\frac{C}{B}\right)^{2}\Bigg[\frac{A''}{A}+\frac{C''}{C}
-\frac{A'}{A}\left(\frac{B'}{B}-\frac{C'}{C}\right)\\ \nonumber
&&-\frac{B'C'}{BC}\Bigg]=\frac{C^{2}}{f_{R}}\bigg[\kappa\left(p_{\perp eff}+2\pi E^{2}\right)-\frac{f-Rf_{R}}{2}+\frac{\ddot{f_{R}}}{A^{2}} -\frac{f_{R}''}{B^{2}}-\frac{\dot{f_{R}}}{A^{2}}\left(\frac{\dot{A}}{A}-\frac{\dot{B}}{B}-\frac{\dot{C}}{C}\right)\\
&&-\frac{f_{R}'}{B^{2}}\left(\frac{A'}{A}-\frac{B'}{B}+\frac{C'}{C}\right)\bigg],
\end{eqnarray}
where $p_{reff}\equiv p_{r}-\frac{4}{\sqrt{3}}\eta \sigma$ and $p_{\perp eff}\equiv p_{\perp}+\frac{2}{\sqrt{3}}\eta \sigma$.
The dynamical equations are very useful for the investigation of gravitational collapse as they provide the energy variation of the collapsing body with time and adjacent surfaces. These equations are also helpful in observing dissipative effects during collapsing process. We can formulate dynamical equations from the contracted Bianchi identities. For this purpose, we use Eq.(\ref{BianchiId}) and get
\begin{eqnarray}\label{15}\nonumber \dot{\rho}+q'\frac{A}{B}+2q\frac{A}{B}\left(\frac{A'}{A}+\frac{C'}{C}\right)+\rho\left(\frac{\dot{B}}{B}+\frac{\dot{C}}{C}\right)+\left(\frac{\dot{B}}{B}+\frac{\dot{C}}{C}\right) \\ +p_{reff}\frac{\dot{B}}{B}+\frac{\dot{B}}{B}+2p_{\perp eff}\frac{\dot{C}}{C}+4\pi E^{2}\left(\frac{\dot{E}}{E}+\frac{\dot{C}}{C}\right)+P_{1}(r,t)=0,
\end{eqnarray}
\begin{eqnarray}\label{16}\nonumber p_{reff}'+p_{reff}\left(\frac{A'}{A}+\frac{C'}{C}\right)+\left(\frac{A'}{A}+\frac{C'}{C}\right)+\rho\frac{A'}{A}+\frac{A'}{A}-2p_{\perp eff}\frac{C'}{C}-2\frac{C'}{C} \\ +\dot{q}\frac{B}{A}+2\frac{B}{A}\left(\frac{\dot{B}}{B}+\frac{\dot{C}}{C}\right)-4\pi E^{2}\left(\frac{E'}{E}+\frac{2C'}{C}\right)+P_{2}(r,t)=0,
\end{eqnarray}
where $P_{1}(r,t)$ and $P_{2}(r,t)$ corresponds to dark source expressions (see appendix).

\section{Perturbation Analysis using Viable $f(R)$ Model}

In this work, we consider the Starobinksy like $f(R)$ model \cite{kau41}
\begin{eqnarray}
f(R)=R+\alpha R^{n}.
\end{eqnarray}
It is mentioned here that for $n=2$, the model reduces to the well known Starobinsky model \cite{star21}.
The viability of model depends upon second order derivative of scalar curvature function. If $f''(R)>0$ then $f(R)$ is assumed to be suitable in general relativity (GR) and Newtonian limits. In proposed $f(R)$ model, $n>2$ and $\alpha$ is positive real number for demonstrating accelerated expansion of the universe and fulfilling stability criterion.
Since the field equations are highly complex and nonlinear, so their solution is difficult to investigate analytically. Therefore theory of perturbation is implemented by assuming that the metric function are static and perturbed quantities are time and radical dependent. Using $0<\epsilon<<1$ , functions can be written as follows.
\begin{eqnarray}A(t,r)=A_{o}(r)+\epsilon T(t)a(r),\end{eqnarray}
\begin{equation}B(t,r)=B_{o}(r)+\epsilon T(t)b(r),\end{equation}
\begin{equation}C(t,r)=C_{o}(r)+\epsilon T(t)\overline{c}(r),\end{equation}
\begin{equation}\rho(t,r)=\rho_{o}(r)+\epsilon \overline{\rho}(t,r),\end{equation}
\begin{equation}p_{reff}(t,r)=p_{roeff}(r)+\epsilon \overline{p}_{reff}(t,r),\end{equation}
\begin{equation}p_{\perp eff}(t,r)=p_{\perp oeff}(r)+\epsilon \overline{p}_{\perp eff}(t,r),\end{equation}
\begin{equation}m(t,r)=m_{o}(r)+\epsilon \overline{m}(t,r),\end{equation}
\begin{equation}q(t,r)=\epsilon \overline{q}(t,r),\end{equation}
\begin{equation}R(t,r)=R_{o}(r)+\epsilon T(t)e(r),\end{equation}
\begin{equation}E(t,r)=E_{o}(r)+\epsilon T(t)h(r),\end{equation}
\begin{equation}f(R)=[R_{o}(r)(1+\alpha R_{o}^{n-1}(r))]+\epsilon T(t)e(r)[1+\alpha nR_{o}^{n-1}(r)],\end{equation}
\begin{equation}f_{R}(R)=1+\alpha nR_{o}^{n-1}(r)+\epsilon\alpha n(n-1)R_{o}^{n-2}(r)T(t)e(r).\end{equation}
Assuming $C_{o}(r)=r$, static configuration of the field equations (\ref{9})-(\ref{14}) takes the form:
\begin{eqnarray}\nonumber
\frac{2B_{o}'}{rB_{o}}-\frac{1}{r^{2}}&=&\frac{\kappa B_{o}^{2}}{1+\alpha nR_{o}^{n-1}}\Bigg[\left(\rho_{o}+1+2\pi E_{o}^{2}\right)+\frac{\alpha n(n-1)R_{o}^{n-2}}{\kappa}\times\\&&\bigg\lbrace-\frac{R_{o}^{2}}{2n}+\frac{(n-2)R_{o}^{-1}}{B_{o}^{2}}-
\frac{1}{B_{o}^{2}}\left(\frac{B_{o}'}{B_{o}}-\frac{2}{r}\right)\bigg\rbrace\Bigg]\\
\nonumber \frac{2A_{o}'}{rA_{o}}+\frac{1}{r^{2}}&=&\frac{\kappa B_{o}^{2}}{1+\alpha nR_{o}^{n-1}}\Bigg[\left(p_{roeff}+1+2\pi E_{o}^{2}\right)+\frac{\alpha n(n-1)R_{o}^{n-2}}{\kappa}\times\\
&&\bigg\lbrace\frac{R_{o}^{2}}{2n}-\frac{1}{B_{o}^{2}}\left(\frac{A_{o}'}{A_{o}}-\frac{2}{r}\right)\bigg\rbrace\Bigg],\end{eqnarray}
\begin{eqnarray}\nonumber \frac{1}{r}\left(\frac{A_{o}'}{A_{o}}-\frac{B_{o}'}{B_{o}}\right)+\frac{A_{o}''}{A_{o}}-\frac{A_{o}'B_{o}'}{A_{o}B_{O}}=\frac{\kappa B_{o}^{2}}{1+\alpha nR_{o}^{n-1}}\Bigg[\frac{\alpha n(n-1)R_{o}^{n-2}}{\kappa}\bigg\lbrace\frac{R_{o}}{2n}\\-\frac{(n-2)R_{o}^{-1}}{B_{o}^{2}}
-\frac{1}{B_{o}^{2}}\left(\frac{A_{o}'}{A_{o}}-\frac{B_{o}'}{B_{o}}+\frac{1}{r}\right)\bigg\rbrace + p_{\perp o eff}+2\pi E_{o}^{2}\Bigg],\end{eqnarray}
First dynamical equation (\ref{15}) is identically satisfied in static configuration, while second evolution equation (\ref{16}) has static configuration as:
\begin{eqnarray} p_{roeff}'+\left(\rho_{o}+p_{roeff}+2\right)\frac{A_{o}'}{A_{o}}+\left(p_{roeff}-p_{\perp oeff}\right)\frac{2}{r}-4\pi E_{o}^{2}\left(\frac{E_{o}'}{E_{o}}+\frac{2}{r}\right)+P_{2s}=0,
\end{eqnarray}
where $P_{2s}$ denotes the static part of $P_{2}(r,t)$ (see appendix).
Perturbed configurations of evolution equations (\ref{15}) and (\ref{16}) are
\begin{eqnarray}\label{17}\nonumber &&\dot{\overline{\rho}}+\overline{q}'\frac{A_{o}}{B_{o}}+2\overline{q}\frac{A_{o}}{B_{o}}\left(\frac{A_{o}'}{A_{o}}+\frac{1}{r}\right)+ \bigg[2\pi E_{o}^{2}\left(\frac{h}{E_{o}}+\frac{2\overline{c}}{r}\right)+\\&&\frac{b}{B_{o}}\left(\rho_{o}+p_{roeff}+2\right)+\frac{2\overline{c}}{r}\left(\rho_{o}+p_{\perp oeff}+3\right)+P_{1p}\bigg]\dot{T}=0,
\end{eqnarray}
\begin{eqnarray}\label{18}\nonumber
&&\overline{p}_{reff}'+\dot{\overline{q}}\frac{B_{o}}{A_{o}}+\left(\overline{\rho}+\overline{p}_{reff}+2\right)\frac{A_{o}'}{A_{o}}+\left(2\overline{p}_{reff}+\overline{p}_{\perp eff}+3\right)\frac{1}{r}+
\\ \nonumber
&&\Bigl[-4\pi\lbrace \left(E_{o}h\right)'+2E_{o}^{2}\left(\frac{\overline{c}}{r}\right)'+2E_{o}h \left(\frac{E_{o}'}{E_{o}}+\frac{2}{r}\right) \rbrace +(\rho_{o}+p_{roeff}+2)\times\\
&& \left(\frac{a}{A_{o}}\right)'+(2p_{roeff}+p_{\perp oeff}+3) \left(\frac{\overline{c}}{r}\right)'\Bigr] T+ P_{2p}=0,\end{eqnarray}
where $P_{1p}$ and $P_{2p}$ denotes perturbed part of $P_{1}$ and $P_{2}$ respectively (see appendix).
Using the perturbation form of equation (\ref{10}), $\overline{q}$  is eliminated as
\begin{eqnarray}\label{20}\nonumber
\overline{q}&=&\frac{1}{\kappa A_{o}B_{o}} \bigg[\alpha n(n-1)R_{o}^{n-2} \Big\{ e'+e(n-2)R_{o}^{-1}R_{o}' -e\frac{A_{o}'}{A_{o}}-\frac{b}{B_{o}}R_{o}'\Big\} \\ &&-2(1+\alpha nR_{o}^{n-1})\Bigg\{\frac{\overline{c}A_{o}'}{rA_{o}}+\frac{b}{rB_{o}}-\frac{\overline{c}'}{r}\Bigg\} \bigg]\dot{T}-1.\end{eqnarray}
Using Eqs.(\ref{17}) and (\ref{20}), we get
\begin{eqnarray}\label{102}
\dot{\overline{\rho}}=\Bigl[-\frac{b}{B_{o}}(\rho_{o}+p_{roeff})-\frac{2\overline{c}}{r}(\rho_{o}+p_{\perp oeff})-4\pi E_{o}^{2}\left(\frac{h}{E_{o}}+\frac{2\overline{c}}{r}\right)+P_{3}(r)\bigr]\dot{T}.\end{eqnarray}
Integrating Eq.(\ref{102}) with respect to $"t"$, it follows
\begin{eqnarray}\label{19}\overline{\rho}=\Bigl[-\frac{b}{B_{o}}(\rho_{o}+p_{roeff})-\frac{2\overline{c}}{r}(\rho_{o}+p_{\perp oeff})-4\pi E_{o}^{2}\left(\frac{h}{E_{o}}+\frac{2\overline{c}}{r}\right)+P_{3}(r)\bigr]T.\end{eqnarray}
The Harrison-Wheeler type equation of state describing second law of ther- modynamics relates $\overline{\rho}$ and $\overline{p}_{reff}$ in terms of adiabatic index $\Gamma$ as \cite{chan10, whe1}

\begin{eqnarray}\overline{p}_{reff}=\Gamma\frac{p_{roeff}}{\rho_{o}+p_{roeff}} \overline{\rho},\end{eqnarray}
The adiabatic index $\Gamma$ is a measure to recognize pressure variation with changing density. Putting equation (\ref{19}) in above equation, we get
\begin{eqnarray}\label{21}\nonumber \overline{p}_{reff}&=&-\Gamma \Bigl[p_{roeff}\frac{b}{B_{o}}+\frac{2\overline{c}}{r}\frac{p_{roeff}(\rho_{o}+p_{\perp oeff})}{\rho_{o}+p_{roeff}}+4\pi E_{o}^{2}\times \\ &&\left(\frac{h}{E_{o}}+\frac{2\overline{c}}{r}\right)\frac{p_{roeff}}{\rho_{o}+p_{roeff}}-\frac{p_{roeff}}{\rho_{o}+p_{roeff}}P_{3}\Bigr] T. \end{eqnarray}
Inserting $\overline{\rho}$, $\overline{q}$ and $\overline{p}_{reff}$ from equations (\ref{19}), (\ref{20}) and (\ref{21}) in (\ref{18}) leads to
\begin{eqnarray}\label{22}\nonumber
&&\frac{\ddot{T}}{\kappa A_{o}^{2}}\bigg[\alpha n(n-1)R_{o}^{n-2}\Bigg\{e'+e(n-2)R_{o}^{-1}R_{o}'-e\frac{A_{o}'}{A_{o}}-\frac{b}{B_{o}}R{o}'\Bigg\} -2(1+\\
\nonumber
&&\alpha nR_{o}^{n-1})\Bigg\{\frac{\overline{c}A_{o}'}{rA_{o}}+\frac{b}{rB_{o}}-\frac{\overline{c}'}{r}\Bigg\} \bigg]-\Gamma T\bigg[p_{roeff}\frac{b}{B_{o}}+\frac{2\overline{c}}{r}\frac{p_{roeff}(\rho_{o}+p_{\perp oeff})}{\rho_{o}+p_{roeff}}\\
\nonumber
&&+\Bigg\{ 4\pi E_{o}^{2}\Bigg[\frac{h}{E_{o}}
+\frac{2\overline{c}}{r}\Bigg]-P_{3}\Bigg\}\frac{p_{roeff}}{\rho_{o}+p_{roeff}}\bigg]_{,1}-\Gamma T\left(\frac{A_{o}'}{A_{o}}+\frac{2}{r}\right)\bigg[p_{roeff}\frac{b}{B_{o}}\\
\nonumber
&&+\frac{2\overline{c}}{r}\frac{p_{roeff}(\rho_{o}+p_{\perp oeff})}{\rho_{o}+p_{roeff}}+
 \Bigg\{ 4\pi E_{o}^{2}\left(\frac{h}{E_{o}}+\frac{2\overline{c}}{r} \right)
-P_{3}\Bigg\}\frac{p_{roeff}}{\rho_{o}+p_{roeff}}\bigg]+\frac{\overline{p}_{\perp eff}}{r}\\
\nonumber
&&-\frac{A_{o}'}{A_{o}}\bigg[-\frac{b}{B_{o}}(\rho_{o}+p_{roeff})-\frac{2\overline{c}}{r}(\rho_{o}+p_{\perp oeff})
-4\pi E_{o}^{2}\left(\frac{h}{E_{o}}+\frac{2\overline{c}}{r}\right)
+P_{3}(r)\bigg]T\\
\nonumber
&&+\bigg[-4\pi\Bigg\{(E_{o}h)'+2E_{o}^{2}\left(\frac{\overline{c}}{r}\right)'+2E_{o}h\left(\frac{E_{o}'}E_{o}+\frac{2}{r}\right)\Bigg\} +(\rho_{o}+p_{roeff})\left(\frac{a}{A_{o}}\right)'\\
&&+(2p_{roeff}+p_{\perp oeff})\left(\frac{\overline{c}}{r}\right)'\bigg] T+P_{2p}=0.\end{eqnarray}
The differential equation obtained from perturb Ricci Scalar curvature is
\begin{eqnarray}\label{222}\ddot{T}(t)-P_{4}(r)T(t)=0.\end{eqnarray}
$P_{4}(r)$ is given in Appendix.
A straight forward solution of above equation is obtained as
\begin{eqnarray}\label{23}T(t)=-e^{\sqrt{P_{4}}t}.\end{eqnarray}
Some other solutions are also possible using separable variable method. However the differential equation involving the radial coordinate $r$ becomes too much complicated to deal with. Therefore we restrict ourselves to consider equation (\ref{23}) and estimate instability range in Newtonian and post-Newtonian regimes.

\subsection{Newtonian Regime}

In this approximation, we assume that $\rho_{o} >> p_{roeff}, \rho_{o} >> p_{\perp oeff}$ and $A_{o}=1, B_{o}=1$.
By substituting these values in equation (\ref{22}), we get
\begin{eqnarray}\nonumber && \frac{\ddot{T}}{\kappa}\Bigl[\alpha n(n-1)R_{o}^{n-2}\lbrace e'+e(n-2)R_{o}^{-1}R_{o}'-bR{o}'\rbrace -\frac{2\left(1+\alpha nR_{o}^{n-1}\right)}{r} \left(b-\overline{c}'\right)\Bigr] \\\nonumber && - \Gamma T\Bigl[p_{roeff}b+\frac{2\overline{c}}{r}p_{roeff}\Bigr]'-\Gamma T \frac{2}{r}\Bigl[p_{roeff}b+\frac{2\overline{c}}{r}p_{roeff}\Bigr]+\frac{\overline{p}_{\perp eff}(N)}{r}+\Bigl[(2p_{roeff}+p_{\perp oeff})\\ && \left(\frac{\overline{c}}{r}\right)'  -4\pi \Big\{(E_{o}h)' +2E_{o}^{2}\left(\frac{\overline{c}}{r}\right)'+2E_{o}h\left(\frac{E_{o}'}{E_{o}}+\frac{2}{r}\right) \Big\} +\rho_{o}a'\Bigr]T+P_{2p}(N)=0. \label{2323}
\end{eqnarray}
Inserting the value of $T$ from Eq.(\ref{23}) in Eq. (\ref{2323}), we obtain
\begin{eqnarray}\nonumber
\Gamma = \frac{1}
{\Bigl(p_{roeff}b+\frac{2\overline{c}}{r}p_{roeff}\Bigr)'+\frac{2p_{roeff}}{r}\Bigl( b+\frac{2\overline{c}}{r}\Bigr)}\bigg[\rho_{o}a'-4\pi\lbrace(E_{o}h)'+2E_{o}^{2}\left(\frac{\overline{c}}{r}\right)'+\\ \nonumber
2E_{o}h(\frac{E_{o}'}{E_{o}}+\frac{2}{r})\rbrace + \frac{\overline{p}_{\perp eff}(N)}{r}+ (2p_{roeff}+p_{\perp oeff})\left(\frac{\overline{c}}{r}\right)'+P_{2p}(N)+P_{5}\bigg].
\end{eqnarray}
Thus the range for instability regime is given by \cite{bha}
\begin{eqnarray}\label{24}\Gamma < \frac{\rho_{o}a'-4\pi\lbrace(E_{o}h)'+2E_{o}^{2}\left(\frac{\overline{c}}{r}\right)'+2E_{o}h(\frac{E_{o}'}{E_{o}}+\frac{2}{r})\rbrace+P_{2p}(N)+P_{5}}
{\Bigl[p_{roeff}b+\frac{2\overline{c}}{r}p_{roeff}\Bigr]'+\frac{2p_{roeff}}{r}\left( b+\frac{2\overline{c}}{r}\right)}.
\end{eqnarray}
This relationship implies that the energy density, the electromagnetic field, the curvature term and the pressure affect the value of the adiabatic index, which in our analysis raises the importance of these physical variables. However, heat flux does not contribute to dynamic instability. It is clear that the range of instability increases with the inclusion of electromagnetic fields. The system approaches the unsteady state of the Newton approximation until the inequality (\ref {24}) holds.

When $\alpha \to 0$, the inequality (\ref{24}) takes the form
\begin{eqnarray}\Gamma < \frac{\rho_{o}a'-4\pi\lbrace(E_{o}h)'+2E_{o}^{2}\left(\frac{\overline{c}}{r}\right)'+2E_{o}h(\frac{E_{o}'}{E_{o}}+\frac{2}{r})\rbrace - \frac{2P_{4}}{r\kappa}\left( b-\overline{c}'\right)}{\Bigl[p_{roeff}b+\frac{2\overline{c}}{r}p_{roeff}\Bigr]'+\frac{2p_{roeff}}{r}\left(b+\frac{2\overline{c}}{r}\right)}.\end{eqnarray}
This corresponds to general relativity.

\subsection{Post Newtonian Regime}

We summarize relativistic impressions upto $O \left(\frac{m_{o}}{r}+\frac{Q^{2}}{2r^{2}}\right)$. In this approximation, we take
\begin{eqnarray}\label{25} A_{o}=1-\frac{m_{o}}{r}+\frac{Q^{2}}{2r^{2}}, ~~~ B_{o}=1+\frac{m_{o}}{r}-\frac{Q^{2}}{2r^{2}},\end{eqnarray}

Using Eq.(\ref{25}) in Eq.(\ref{22}), we get
\begin{eqnarray}\Gamma < \frac{W+X+\frac{\overline{p}_{\perp eff (PN)}}{r}+P_{2(PN)}}{N'-\frac{2}{r}\frac{rm_{o}-2r^{2}}{2rm_{o}-2r^{2}-Q^{2}}N},\label{2313}
\end{eqnarray}
where $W$, $X$ and $N$ forms expressions as,
\begin{eqnarray}\nonumber
W&=&\frac{4r^{4}P_{4(PN)}}{\kappa (2r^{2}+2rm_{o}-Q^{2})^{2}}\Big[\alpha n(n-1)R_{o}^{n-2} \Big\{ e'+e(n-2)R_{o}^{-1}R_{o}'\Big\}\\\nonumber
&&-\frac{2}{r}\frac{Q^{2}-rm_{o}}{2rm_{o}-2r^{2}-Q^{2}}\Big\{ e\alpha n(n-1)R_{o}^{n-2}+\frac{2\overline{c}}{r}\left(1+\alpha nR_{o}^{n-1}\right)\Big\}\\\nonumber
&&-\frac{2r^{2}}{2r^{2}+2rm_{o}-Q^{2}}\Big\{ b\alpha n(n-1)R_{o}^{n-2}R_{o}'+\frac{2b}{r}\left(1+\alpha nR_{o}^{n-1}\right)\Big\}\\
&&+\frac{2\overline{c}'}{r}\left(1+\alpha nR_{o}^{n-1}\right)\Big],\end{eqnarray}
\begin{eqnarray}\nonumber X&=&-\frac{2}{r}\frac{Q^{2}-rm_{o}}{2rm_{o}-2r^{2}-Q^{2}}\Bigg[\frac{2br^{2}(\rho_{o}+p_{reff})}{2rm_{o}+2r^{2}-Q^{2}}-\frac{2\overline{c}}{r}(\rho_{o}+p_{\perp eff})+P_{4(PN)}\\\nonumber &&-4\pi E_{o}^{2}\left(\frac{h}{E_{o}}+\frac{2\overline{c}}{r}\right)\Bigg]+4\pi \Bigg\{ (E_{o}h)'+2E_{o}^{2}\left(\frac{\overline{c}}{r}\right)'+2E_{o}h\left(\frac{E_{o}'}{E_{o}}+\frac{2}{r}\right) \Bigg\} \\&&-(\rho_{o}+p_{reff})\left(\frac{2ar^{2}}{2rm_{o}-2r^{2}+Q^{2}}\right)'+(2p_{reff}+p_{\perp eff})\left(\frac{\overline{c}}{r}\right)',\end{eqnarray}
\begin{eqnarray}\nonumber
N&=&\frac{2br^{2}p_{reff}}{2rm_{o}+2r^{2}-Q^{2}}+\frac{p_{reff}}{\rho_{o}+p_{reff}}\Bigl[4\pi E_{o}^{2}\left(\frac{h}{E_{o}}+\frac{2\overline{c}}{r}\right)-P_{4(PN)}\Bigr]\\
&&+\frac{2\overline{c}}{r}\frac{p_{reff}(\rho_{o}+{p_{\perp eff}})}{\rho_{o}+p_{reff}}.\end{eqnarray}
The dynamical instability in PN limit remains valid in the region where inequality (\ref{2313}) is satisfied.
In asymptotic limit, the inequality (\ref{2313}) remains the same, however, $W$ takes the form
\begin{eqnarray}W=\frac{4r^{4}}{\kappa (2rm_{o}+2r^{2}-Q^{2})^{2}(2rm_{o}-2r^{2}-Q^{2})}\Bigl[\frac{2\overline{c}}{r^{2}}(Q^{2}-rm_{o})-4br+\frac{2\overline{c}'}{r}\Bigr].
\end{eqnarray}

\section{Summary and Conclusion}

The purpose of this work is to investigate dissipative cylindrical collapse of a charged anisotropic fluid in $f(R)$ Gravity. We consider the $f(R)$ model $f(R)=R+ \alpha R^{n}$ \cite{kau41} and apply perturbation scheme to the modified field equations. It is mentioned here that for $n=2$, the model reduces to the well known Starobinsky model \cite{star21}. The model under consideration provides a viable alternative to dark energy problems and satisfies the condition $f''(R)> 0$. The dynamical equations have been developed to study the evolution of cylindrical stars.  Dissipation in terms of heat flow plays an important role in the dynamics of collapse. In particular the charge and its distribution imply drastic effects on evolution of the stellar structure.

Perturbed form of second Bianchi identity describes the evolution of the collapsing system and is further used to establish the adiabatic index $\Gamma$ which gives the region of instability. The results are analyzed both in Newtonian and Post Newtonian Regimes. It is clear that $\Gamma$ has a dependence on the electric field strength, the radiation effect, the density and the pressure configuration. It is mentioned here that the results reduce to GR in the limiting case when $\alpha \to 0$. Moreover, our results agree with the previous work \cite{sh2} in the case when fluid is non-viscous with isotropic pressure. In the absence of Maxwell source results correspond to the work presented in \cite{sh33}.
When we take $n = 2$ in the model, the results support the parameters in \cite{sh34}.

\section*{Appendix}
\begin{eqnarray}\nonumber
&&P_{1}(r,t)=\frac{1}{\kappa} \Bigg[A^{2} \Bigg\{ \frac{1}{A^{2}}\left(\frac{f-Rf_{R}}{2}-\frac{\dot{f}_{R}}{A^{2}}\left(\frac{\dot{B}}{B}+\frac{2\dot{C}}{C}\right)-\frac{f'_{R}}{B^{2}}\left(\frac{B'}{B}-\frac{2C'}{C}\right)+\frac{f''_{R}}{B^{2}}\right)\Bigg\}  _{,0}\\&& \nonumber +A^{2} \Bigg\{ \frac{1}{A^{2}B^{2}}\left(\dot{f}'_{R}-\frac{A'}{A}\dot{f}_{R}-\frac{\dot{B}}{B}f'_{R}\right)\Bigg\}  _{,1}-\frac{\dot{f}_{R}}{A^{2}}\Bigg\{ \left(\frac{3A'}{A}+\frac{B'}{B}+\frac{2C'}{C}\right)\frac{AA'}{B^{2}}+\left(\frac{\dot{B}}{B}\right)^{2}\\&& \nonumber +2\left(\frac{\dot{C}}{C}\right)^{2}+\frac{3\dot{A}}{A}\left(\frac{\dot{B}}{B}+\frac{2\dot{C}}{C}\right)\Bigg\}+\frac{\dot{f}'_{R}}{B^{2}}\left(\frac{3A'}{A}+\frac{B'}{B}+\frac{2C'}{C}\right)-\frac{2f'_{R}}{B^{2}}\Bigg\{\frac{A'}{A}\left(\frac{2\dot{B}}{B}+\frac{\dot{C}}{C}\right)\\ \nonumber &&+\frac{B'}{B}\left(\frac{\dot{A}}{A}+\frac{\dot{B}}{B}\right)-\frac{C'}{C}\left(\frac{2\dot{A}}{A}-\frac{\dot{B}}{B}+\frac{\dot{C}}{C}\right)\Bigg\}+\frac{f''_{R}}{B^{2}}\left(\frac{2\dot{A}}{A}+\frac{\dot{B}}{B}\right)+\frac{\dot{A}}{A}\left(f-Rf_{R}\right)\\ &&+\frac{\ddot{f}_{R}}{A^{2}}\left(\frac{\dot{B}}{B}+\frac{2\dot{C}}{C}\right) \Bigg],\end{eqnarray}

\begin{eqnarray}\nonumber
&&P_{2}(r,t)=\frac{1}{\kappa} \Bigg[B^{2} \Bigg\{ \frac{1}{B^{2}}\left(\frac{Rf_{R}-f}{2}-\frac{\dot{f}_{R}}{A^{2}}\left(\frac{\dot{A}}{A}-\frac{2\dot{C}}{C}\right)-\frac{f'_{R}}{B^{2}}\left(\frac{A'}{A}+\frac{2C'}{C}\right)+\frac{\ddot{f}_{R}}{A^{2}}\right)\Bigg\}  _{,1}\\
\nonumber &&+B^{2}\Bigg\{\frac{1}{A^{2}B^{2}}\left(\dot{f}'_{R}-\frac{A'}{A}\dot{f}_{R}-\frac{\dot{B}}{B}f'_{R}\right)\Bigg\}_{,0}+\frac{A'}{A}\Bigg\{\frac{\ddot{f}_{R}}{A^{2}}+\frac{f''_{R}}{B^{2}}-\frac{\dot{f}_{R}}{A^{2}}\left(\frac{\dot{A}}{A}+\frac{\dot{B}}{B}\right)\\ \nonumber &&-\frac{f'_{R}}{B^{2}}\left(\frac{A'}{A}+\frac{B'}{B}\right)\Bigg\}+\frac{2B'}{B}\Bigg\{\frac{Rf_{R}-f}{2}+\frac{\ddot{f}_{R}}{A^{2}}-\frac{\dot{f}_{R}}{A^{2}}\left(\frac{\dot{A}}{A}-\frac{2\dot{C}}{C}\right)-\frac{f'_{R}}{B^{2}}\\ \nonumber &&\left(\frac{A'}{A}+\frac{3C'}{C}\right)\Bigg\}+\frac{1}{A^{2}}\left(\frac{\dot{A}}{A}+\frac{3\dot{B}}{B}+\frac{2\dot{C}}{C}\right)\left(\dot{f}'_{R}-\frac{A'}{A}\dot{f}_{R}-\frac{\dot{B}}{B}f'_{R}\right)+\frac{2C'}{C} \Bigg\{\frac{f''_{R}}{B^{2}}\\&&+\frac{\dot{f}_{R}}{A^{2}}\left(\frac{\dot{C}}{C}-\frac{2\dot{B}}{B}\right)-\frac{f'_{R}}{B^{2}}\frac{C'}{C}\Bigg\} \Bigg].\end{eqnarray}

\begin{eqnarray}\nonumber
P_{2s}&=& \frac{\alpha n(n-1)}{\kappa}\Bigg[B^{2}_{o}\Bigg\{\frac{R^{n-2}_{o}}{nB^{2}_{o}}\left(\frac{R^{2}_{o}}{2}-\frac{nR'_{o}}{B^{2}_{o}}\left(\frac{A'_{o}}{A_{o}}+\frac{2}{r}\right)\right)\Bigg\}_{,1}+\left(\frac{A'_{o}}{A_{o}}+\frac{2}{r}\right)\\ \nonumber
&&\left(\frac{R''_{o}+(n-2)R^{-1}_{o}R'^{2}_{o}}{B^{2}_{o}}\right)+\frac{2B'_{o}R^{2}_{o}}{2nB_{o}}\frac{R^{n-2}_{o}R'_{o}}{B^{2}_{o}}\Bigg\{\frac{A'_{o}}{A_{o}}\left(\frac{A'_{o}}{A_{o}}+\frac{B'_{o}}{B_{o}}\right)+\\
&&\frac{2B'_{o}}{B_{o}}\left(\frac{A'_{o}}{A_{o}}+\frac{3}{r}\right)+\frac{2}{r^{2}}\Bigg\}\Bigg].\end{eqnarray}

\begin{eqnarray}\nonumber
P_{1p}&=&\frac{e''+\alpha n(n-1)R^{n-2}_{o}}{\kappa B^{2}_{o}}\Big[(n-2)\Bigg\{(2e'-e)R^{-1}_{o}R'_{o}+eR^{-1}_{o}R''_{o}+e(n-3)R^{-2}_{o}R'^{2}_{o}\Bigg\}\\ \nonumber &&-\frac{R_{o}B^{2}_{o}}{2}-\frac{b}{B_{o}}\Big(R''_{o}+(n-2)R^{-1}_{o}R'^{2}_{o}\Big)+e'+R'_{o}\Bigg\{\frac{\overline{c}}{r^{2}}+\frac{b'}{B_{o}}+\frac{2\overline{c}'}{r}-\frac{b}{B_{o}}\Bigg\{\frac{2A'_{o}}{A_{o}}+\frac{3B'_{o}}{B_{o}}\\ \nonumber &&+\frac{4}{r}-1\Bigg\}-\frac{2\overline{c}}{r}\left(\frac{A'_{o}}{A_{o}}-\frac{2}{r}\right)\Bigg\}+\left(e'-e\frac{A'_{o}}{A_{o}}+(n-2)eR^{-1}_{o}R'_{o}\right) \left(\frac{3A'_{o}}{A_{o}}+\frac{B'_{o}}{B_{o}}+\frac{1}{r}\right)\Bigg]\\&& +\frac{\alpha n(n-1)}{\kappa}\frac{A^{2}_{o}}{B^{2}_{o}}\Bigg[\frac{R^{n-2}_{o}}{A^{2}_{o}B^{2}_{o}}\Bigg\{e'-e\frac{A'_{o}}{A_{o}}-\frac{b}{B_{o}}R'_{o}+(n-2)eR^{-1}_{o}R'_{o}\Bigg\}\Bigg]_{,1}.\end{eqnarray}

\begin{eqnarray}\nonumber p_{2p}&=&\frac{\alpha n(n-1)}{\kappa}\Bigg[\frac{R^{n-2}_{o}}{A^{2}_{o}}\ddot{T}\Bigg\{\left(e'+(n-2)eR^{-1}_{o}R'_{o}\right)\left(1+A^{2}_{o}\right)-\frac{b}{B_{o}}R'_{o}+2e(1-A^{2}_{o})\frac{B'_{o}}{B_{o}}\Bigg\}\\ \nonumber &&+2TbB_{o}\Bigg\{\frac{R^{n-2}_{o}}{B^{4}_{o}}\left(R'_{o}\left(\frac{A'_{o}}{A_{o}}+\frac{2}{r}\right)-\frac{R^{2}_{o}B^{2}_{o}}{2n}\right)\Bigg\}_{,1}+TB^{2}_{o}\Bigg\{\frac{R^{n-2}_{o}}{B^{4}_{o}}\Bigg[e\frac{R_{o}B^{2}_{o}}{2}+\left(\frac{A'_{o}}{A_{o}}+\frac{2}{r}\right)\\ \nonumber &&\Bigg\{e(n-2)R^{-1}_{o}R'_{o}+\frac{R^{2}_{o}bB_{o}}{n}-R'_{o}\Bigg[\left(\frac{a}{A_{o}}\right)'+2\left(\frac{\overline{c}}{r}\right)'+\frac{4b}{B_{o}}\Bigg]+e'\Bigg\}\Bigg]\Bigg\}_{,1}+\frac{R^{n-2}_{o}}{B^{2}_{o}}T \\ \nonumber &&\Bigg\{\left(\frac{A'_{o}}{A_{o}}+\frac{2}{r}\right)\Bigg\{e''+\Bigg[2e'R^{-1}_{o}R'{o}+eR^{-1}_{o}R''_{o}+e(n-3)R^{-2}_{o}R'^{2}_{o}\Bigg](n-2)\Bigg\}\\ \nonumber &&-\Bigg\{2\left(\frac{a}{A_{o}}\right)'\left(\frac{\overline{c}}{r}\right)'+\frac{2b}{B_{o}}\left(\frac{A'_{o}}{A_{o}}+\frac{2}{r}\right)\Bigg\}\Bigg[R''_{o}+(n-2)R^{-1}_{o}R'^{2}_{o}\Bigg]+R'_{o}\Bigg\{\left(\frac{a}{A_{o}}\right)'\\ \nonumber &&\left(\frac{2A'_{o}}{A_{o}}+\frac{3B'_{o}}{B_{o}}\right)+3\left(\frac{b}{B_{o}}\right)'\left(\frac{A'_{o}}{A_{o}}+\frac{2}{r}\right)+2\left(\frac{\overline{c}}{r}\right)'\left(\frac{3B'_{o}}{B_{o}}+\frac{2}{r}\right)\Bigg\}+\Bigg[e'\\ &&+e(n-2)R^{-1}_{o}R'_{o}-\frac{2b}{B_{o}}R'_{o}\Bigg]\Bigg[\frac{A'_{o}}{A_{o}}\left(\frac{A'_{o}}{A_{o}}+\frac{B'_{o}}{B_{o}}\right)+\frac{2}{r}\left(\frac{3B'_{o}}{B_{o}}+\frac{1}{r}\right)\Bigg]\Bigg\} \Bigg].\end{eqnarray}

\begin{eqnarray}\nonumber
P_{3}&=&-\frac{A_{o}}{B_{o}}\Bigg[\frac{1}{\kappa A_{o}B_{o}}\Bigg\{\alpha n(n-1)R^{n-2}_{o}\left(e'+e(n-2)R^{-1}_{o}R'_{o}-e\frac{A'_{o}}{A_{o}}-\frac{b}{B_{o}}R'_{o}\right)\\\nonumber
&&-2\left(1+\alpha nR^{n-1}_{o}\right)\left(\frac{\overline{c}A'_{o}}{rA_{o}}+\frac{b}{rB_{o}}-\frac{\overline{c}'}{r}\right)\Bigg\}\Bigg]_{,1}-\frac{2}{\kappa B^{2}_{o}}\Bigg[\alpha n(n-1)R^{n-2}_{o}\\\nonumber
&&\left(e'+e(n-2)R^{-1}_{o}R'_{o}-e\frac{A'_{o}}{A_{o}}-\frac{b}{B_{o}}R'_{o}\right)-2\left(1+\alpha nR^{n-1}_{o}\right)\Bigg\{\frac{\overline{c}A'_{o}}{rA_{o}}\\
&& +\frac{b}{rB_{o}}-\frac{\overline{c}'}{r}\Bigg\}\Bigg] \left(\frac{A'_{o}}{A_{o}}+\frac{1}{r}\right)-P_{1p}.\end{eqnarray}

\begin{eqnarray}\nonumber P_{4}&=&-\frac{rA^{2}_{o}B_{o}}{br+2B_{o}\overline{c}}\Bigg[\frac{e}{2}-\frac{2\overline{c}}{r^{3}}-\frac{1}{A_{o}B^{2}_{o}}\Bigg\{A''_{o}\Bigg[\frac{a}{A_{o}}+\frac{2b}{B_{o}}\Bigg]-\frac{1}{B_{o}}\Bigg[a'B'_{o}+a''+A'_{o}b'\\ \nonumber &&-A'_{o}B'_{o}\left(\frac{a}{A_{o}}+\frac{3b}{B_{o}}\right)\Bigg]+\frac{2}{r}\Bigg\{a'+\overline{c}'A'_{o}-A'_{o}\Bigg[\frac{a}{A_{o}}+\frac{2b}{B_{o}}+\frac{\overline{c}}{r}\Bigg]\Bigg\}+\\ &&\frac{A_{o}}{r}\Bigg\{\overline{c}''-\frac{b'}{B_{o}} -\frac{B'_{o}\overline{c}'}{B_{o}}+\frac{3b}{B_{o}}+\frac{\overline{c}}{r}\Bigg\}+\frac{2}{r^{2}}\Bigg[\overline{c}'-\frac{b}{B_{o}}-\frac{\overline{c}}{r}\Bigg]\Bigg\}\Bigg].\end{eqnarray}

\begin{eqnarray}P_{5}=\frac{P_{4}}{\kappa}\Bigg[\alpha n(n-1)R^{n-2}_{o}\Bigg\{e'+e(n-2)R^{-1}_{o}-bR'_{o}\Bigg\}-\frac{2(1+\alpha nR^{n-1}_{o})}{r}\left( b-\overline{c}'\right) \Bigg].\end{eqnarray}

\end{document}